\documentclass[a4paper]{article}

\usepackage{INTERSPEECH2020}

\usepackage{color}
\usepackage{diagbox}
\usepackage{xcolor,colortbl}
\usepackage{enumitem}
\usepackage{amssymb}
\usepackage{balance}

\usepackage{times}
\usepackage{graphicx}
\usepackage{amssymb}
\usepackage{gensymb}
\usepackage{amsmath}
\usepackage{breakurl}
\usepackage{caption}
\usepackage{subcaption}
\usepackage{tikz}
\usetikzlibrary{shapes.geometric, positioning, arrows, bayesnet}

\usepackage{url,hyperref}

\usepackage{algorithm}
\usepackage{algorithmic}

\usepackage[labelfont={bf},font=small]{caption}
\usepackage[none]{hyphenat}

\usepackage{mathtools, cuted}

\usepackage[noadjust, nobreak]{cite}

\usepackage{tabularx}
\usepackage{amsmath}

\usepackage{float}

\usepackage{pifont}

\newcolumntype{Y}{>{\centering\arraybackslash}X}

\usepackage[]{placeins}

\usepackage{placeins}

\usepackage{tikz}

\usepackage[framemethod=tikz]{mdframed}

\usepackage{afterpage}

\usepackage{stfloats}

\usepackage{atbegshi}
\newcommand{\handlethispage}{}
\newcommand{\discardpagesfromhere}{\let\handlethispage\AtBeginShipoutDiscard}
\newcommand{\keeppagesfromhere}{\let\handlethispage\relax}
\AtBeginShipout{\handlethispage}

\usepackage{comment}

\usepackage[activate]{microtype}
\title{On Deep Speech Packet Loss Concealment: A Mini-Survey}

\name{Mostafa M. Mohamed$^{1,2}$, Mina A.\ Nessiem$^{1,2}$, and Bj\"orn W.\ Schuller$^{2,3}$}
\address{
  $^1$Chair of Embedded Intelligence for Health Care and Wellbeing, University of Augsburg, Germany\\
  $^2$AI R\&D team, SyncPilot GmbH, Augsburg, Germany\\
  $^3$GLAM -- Group on Language, Audio, \& Music, Imperial College London, UK}
\email{mostafa.amin@syncpilot.com, mina.nessiem@syncpilot.com, schuller@IEEE.org}

\begin{document}

\maketitle
\thispagestyle{empty}
\pagestyle{empty}

\begin{abstract}
Packet-loss is a common problem in data transmission, using Voice over IP. The problem is an old problem, and there has been a variety of classical approaches that were developed to overcome this problem. However, with the rise of deep learning and generative models like Generative Adversarial Networks and 
Autoencoders, a new avenue has emerged for attempting to solve packet-loss using deep learning, by generating replacements for lost packets. In this mini-survey, we review all the literature we found to date, that attempt to solve the packet-loss in speech using deep learning methods. Additionally, we briefly review how the problem of packet-loss in a realistic setting is modelled, and how to evaluate Packet Loss Concealment techniques. Moreover, we review a few modern deep learning techniques in related domains that have shown promising results. These techniques shed light on future potentially better solutions for PLC and additional challenges that need to be considered simultaneously with packet-loss.

\end{abstract}


\section{Introduction}

Packet-loss is a common problem in data transmission using Voice over IP (VoIP), where audio data is transmitted by first dividing it into small chunks (or \textit{packets}) before sending. During packet transmission, a variety of issues can occur, such as packet delay, packet-loss, or jitter \cite{qos}. When it comes to audio transmission in particular, packet-loss results in the loss of audio content, while packet delay results in meaningless audio playback. In addition to these, other problems contribute to audio quality degradation, such as echo and coded speech \cite{qos}.

Packet Loss Concealment (PLC) is any technique that attempts to overcome the packet-loss problem, by concealing the lost fragments by some estimated reconstruction of the lost audio. Ideally, it would fully restore the lost audio fragments. Basic PLC techniques are 0s filling in place of the lost fragments, repetition of the segments before the loss, or interpolation based methods \cite{perkins1998survey}. More sophisticated techniques are utilised to solve the issue, like Linear Predictive Coding (LPC) \cite{LPCoriginal} which is adapted in GSM technologies \cite{GSM}. Other approaches have attempted to solve the problem based on speech coding methods, for example, sending extra low-rate redundant packet after an original packet that could serve as a replacement in case of loss \cite{codecs}. PLC techniques can be seen as sender-based or receiver-based \cite{perkins1998survey}, in addition to a hybrid between both \cite{codecs}. 

In the age of deep learning, there has been a variety of generative models that have been developed, like Generative Adversarial Networks (GANs) \cite{GANS}, sequential generation using Recurrent Neural Networks (RNN) \cite{graves} and Autoencoders \cite{autoencoder}. These models have proven superior to generate or fix data in the domains of text \cite{graves} and images \cite{nvidia}. There are also attempts to use such models for audio generation \cite{wavegan}. Hence, this has motivated several approaches to solve PLC using models based on neural networks. Typically, existing deep PLC approaches are receiver-based, where speech is post-processed after the packet-loss to generate concealed packets using the deep models. 
There are older surveys already reviewing this problem \cite{perkins1998survey, voipsurvey}, however, to the best of our knowledge, there are no surveys dealing with modern approaches based on deep learning. Hence, the aim of this paper is to provide a mini-survey to review all the modern deep PLC approaches for speech and shed light on potential future solutions and challenges.

This paper is divided as follows: in Section \ref{sec:modelling}, we will review how the problem is modelled in realistic settings, in addition to some evaluation techniques and a brief review of the classical PLC techniques. The deep speech PLC approaches will be reviewed in Section \ref{sec:approaches}. In Section \ref{sec:challenges}, we will discuss some challenges and potential future directions for deep PLC. Eventually, we summarise the paper in Section \ref{sec:conclusion}.

\section{Modelling and Background}
\label{sec:modelling}

In order to experiment with Packet Loss Concealment (PLC) techniques, there needs to be a model that could simulate the packet-loss behaviour. Such modellings are discussed in Subsection \ref{subsec:markovs}. After that, evaluation methods are needed to evaluate the performance. Such methods are discussed in Subsection \ref{subsec:evals}. Finally, a background of the classical PLC techniques is given in Subsection \ref{subsec:classical}.

\begin{figure}[h!]
    \centering
    \begin{tikzpicture}
        
        \node[circle, draw] (qn) {$N:1$};
        \node[] at (-1, 0) (start) {};
        \node[circle, draw] at (3, 0) (ql) {$L:0$};
        
        \draw 
        (start) [->] edge (qn)
        (qn) edge[loop above] node{$p_{\text{N}}$} (qn)
        (ql)[->] edge[loop above] node{$p_{\text{L}}$} (ql)
        (qn)[->] edge[bend left, above] node{$1 - p_{\text{N}}$} (ql)
        (ql) edge[bend left, below] node{$1 - p_{\text{L}}$} (qn);
    \end{tikzpicture}

    \caption{Two-state Markov model for modelling packet loss. The states represent \textit{loss} and \textit{non-loss} scenarios. This can sample loss/non-loss binary masks.}
    \label{fig:markov}
    \vspace{-0.4cm}
\end{figure}
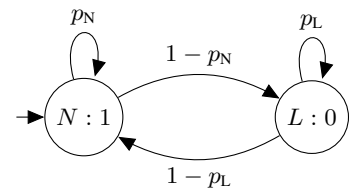

\subsection{Modelling packet-loss behaviour}
\label{subsec:markovs}
Modelling packet-loss in a real-time setting is essential to be able to solve the packet-loss problem. PLC techniques need to be validated in realistic scenarios in order to be successful. Several models are developed for this, including two-state Markov models (depicted in Figure \ref{fig:markov}), Gilbert-Elliot models,  
or three-state Markov models \cite{da2019mac}. 
A modern survey \cite{da2019mac} is reviewing several of these approaches with a detailed comparison.

\subsection{Evaluation methods}
\label{subsec:evals}
Evaluating a given PLC mechanism is not a straight forward issue to handle. This evaluation can be categorised as an instance of Speech Enhancement (SE) evaluation. The main reason behind this difficulty is that, there are several factors involved in the perceived quality of speech. A variety of measures capture some of these factors, however, they still suffer from not fully modelling realistic factors of speech deformation \cite{SpeechEnhancementBook}.
Mean Opinion Score (MOS) is a common way for evaluation, where listeners manually annotate the enhancement on a five-degree scale. Then, the annotated ratings are averaged \cite{SpeechEnhancementBook}. 
Furthermore, there was a competition held in 2002 to develop an objective metric for SE tasks. Perceptual Evaluation of Speech Quality measure (PESQ) was chosen \cite{SpeechEnhancementBook}.
One of the semantic factors is the intelligibility of speech. Short-Time Objective Intelligibility measure (STOI) \cite{STOI} is an evaluation method that addresses this factor. 

Moreover, there are objective measures for SE tasks. There is a comprehensive chapter in \cite{SpeechEnhancementBook} that describes many of them.
Standard metrics can be used as objective measures, for example, \textit{Mean Squared Error} (MSE), \textit{Mean Absolute Error} (MAE), \textit{Mean Absolute Percentage Error} (MAPE), and Pearson's correlation coefficient \cite{schuller2013}.
Concordance Correlation Coefficient (CCC) \cite{ccc} measures data reproducibility, which is trending in Speech Emotion Recognition \cite{tzirakis2018}. Additional sound specific measures are: Log Spectral Distance (LSD) \cite{LSD}, 
 Signal to Noise Ratio (SNR), Signal to Interference Ratio (SIR), Signal to Distortion Ratio (SDR), and Signal to Artifacts Ratio (SAR) \cite{vincent2006performance}.

\subsection{Classical approaches}
\label{subsec:classical}

There are various classical techniques for PLC -- we mention a few highlights here.
Linear Predictive Coding (LPC) \cite{LPCoriginal} is a widely used technique in GSM technologies \cite{GSM}. It is based on predicting a value $\hat{x}_t$ for a lost frame from few preceding $k$ frames $x_{t - k}, \cdots , x_{t-1}$, according to the linear recurrence:
$$    \hat{x}_t = \sum_{i=1}^k a_i x_{t - i}$$

One statistical-based technique is Hidden Markov Models (HMMs)  \cite{rodbro2006hidden, hmminterspeech}, where the preceding packets are also used to estimate the missing packet. However, the prediction is based on HMMs instead of linear recurrence as in LPC.  
An autoregressive model is used in \cite{autoregressive}, which is similar to LPC with an additional incorporation of a stochastic random variable drawn from a Gaussian distribution. A later approach \cite{interspeech-sigmoid} incorporated an adaptive model with a sigmoid output function for muting.
Another class of approaches is based on speech codings that are designed to be robust against packet-loss \cite{codecs, sinusoidal_codec}.



\section{Deep Approaches}
\label{sec:approaches}

There are two general frameworks how Packet Loss Concealment (PLC) operates. The first is in real-time settings, where upon receiving each segment, it is post-processed using the PLC method (potentially in addition to other post-processing). The post-processed frame is either the original non-lost frame or the concealed lost frame. Typically, the segments are of short length mirroring realistic packet sizes (10-20\,ms) \cite{qos}. 
The second framework is processing larger segments of audio including some lost packets, using deep generative models like Generative Adversarial Networks (GANs) or Autoencoders. Overall, the non-lost parts provide a general context, based on which, the PLC techniques can conceal the lost segments. The approaches of the second type are typical for offline processing because of using broader contexts that might include future segments, in addition to the large segment processing and usage of deep models, which makes them not the most suitable for real-time fast processing. The real-time approaches are described in Subsection \ref{subsec:realtime}, the Autoencoders approaches are discussed in Subsection \ref{subsec:autoencoders}, and the GAN approaches are described in Subsection \ref{subsec:GAN}. 
Moreover, the approaches are listed in Table \ref{tab:approaches}.

\begin{figure}[!t]
    \centering
    \begin{tikzpicture}
    \node[rectangle, draw, rounded corners, minimum height=0.5cm, minimum width=2cm] at (0, 4.5) (frame) {input}; 
    \node[diamond, draw, minimum height=0.5cm, minimum width=2cm] at (0, 3.2) (lost) {loss ?}; 
    \node[rectangle, draw, minimum height=0.5cm, minimum width=2cm] at (0, 2) (history) {history}; 
    \node[rectangle, draw, minimum height=0.5cm, minimum width=2cm] at (3.2, 2) (model) {deep conceal}; 
    \node[rectangle, draw, rounded corners, minimum height=0.5cm, minimum width=2cm] at (0, 0.5) (output) {output}; 
    \node[rectangle, draw, minimum height=0.5cm, minimum width=2cm] at (-3, 2) (post) {post-process}; 
    \node[rectangle, draw, minimum height=0.5cm, minimum width=2cm] at (3.2, 0.5) (invert) {invert};
    
    \draw (frame) [->] edge (lost);
    \draw (lost) [-, above] edge node{Yes} (3.2, 3.2);
    \draw  [->] (3.2, 3.2) --  (model);
    
    \draw (lost) [-, above] edge node{No} (-3, 3.2);
    \draw  [->] (-3, 3.2) --  (post);
    
    \draw (post) [->] edge (history);
    \draw (history) [->] edge (model);
    
    \draw (invert) [->] -- (output);

    
    \draw (post) [-] edge (-3, 0.5);
    \draw  (-3, 0.5) [->] --  (output);
    
    \draw (3.2, 1.3) [-] -- (0, 1.3);
    \draw (0, 1.3) [->] -- (history);
    
    \draw (model) [->] edge (invert);
    
    \end{tikzpicture}

    \caption{Underlying principle for real-time PLC algorithms. For each non-lost frame, it can be post-processed then it goes to the output, otherwise the deep model estimates the features of the lost frame and inverts them to audio content. A history context is maintained for future predictions.}
    \vspace{-0.4cm}
    \label{fig:PLC}
\end{figure}
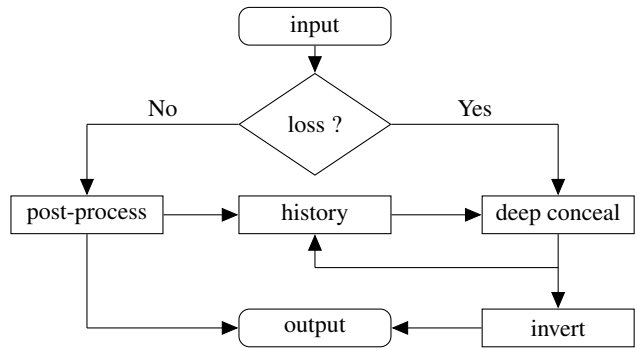

\subsection{Real-time deep PLC approaches}
\label{subsec:realtime}
To the best of our knowledge, \cite{PLC2016} is the first paper that addresses the packet-loss problem using deep learning. They introduce a framework for a PLC algorithm -- its underlying principle is depicted in Figure \ref{fig:PLC}. The authors deal with the problem on the feature level of speech. First, by training a feed-forward neural network, given the features of $P$ consecutive frames, the network predicts the features of the succeeding frame. In the inference phase, for each frame, if it is not lost, then, it is decoded as it is; otherwise, the network predicts a concealment for the features of the lost frame, then, its inverse is generated which corresponds to the predicted lost audio.

One of the characteristics of this approach is that it relies on invertible features. Consequently, the presented PLC framework is valid for any features that have this property. However, this offers a limitation, which is the need for feature selection; the inversion adds a restriction on the possible features, which could be sub-optimal for some speech tasks.
Another limitation of this approach is using a fixed length context of $P$ preceding frames to estimate the lost frame. This neglects potentially useful information that are maintained over long-term dependencies. 

Both of these limitations were overcame in \cite{Lotfidereshgi2018}, while maintaining the same framework; they used LSTM \cite{LSTM} on raw audio frames directly to estimate the succeeding frame. One of the main advantages of using gated recurrent cells like LSTM \cite{LSTM} 
is handling long-term dependencies in sequential data,
by maintaining internal memories. This has been shown to be very effective in language modelling as well \cite{graves}. 

A new trend in deep learning is to use end-to-end networks for a variety of tasks like Automatic Speech Recognition (ASR) \cite{ASR} and Speech Emotion Recognition (SER) \cite{tzirakis2018}. One crucial advantage of \cite{Lotfidereshgi2018} is using the raw audio. This removes the need for feature selection and their need to be invertible. On the other hand, for applications that need specific invertible features, still the framework in \cite{PLC2016} will be useful.

An additional crucial advantage of \cite{Lotfidereshgi2018} is adapting online-training into the framework, where every frame that is not lost, can be used to enhance the model by making a training step. This would make the trained model more enhanced when it is used for a long time. A similar idea was earlier employed in the classical approach \cite{interspeech-sigmoid}, using a sigmoid model, which can be thought of as a shallow version of \cite{Lotfidereshgi2018}. 

In an earlier work of the authors \cite{mywork}, an encapsulation of the framework in Figure \ref{fig:PLC} was introduced, where a Recurrent Neural Network (RNN) executes this whole frame end-to-end at inference time.

\subsection{Autoencoder-based deep PLC approaches}
\label{subsec:autoencoders}

The authors in \cite{marafioti2019context, marafioti2019audio} turn to an encoder-decoder architecture in which the inputs are encoded into a lower-dimensional representation that is then used to reconstruct the signal. Rather than dealing with the problem in the time domain, the authors transform the audio into its Time-Frequency (TF) spectrogram representation, which can be thought of as an image, via the Short-Term Fourier Transform (STFT) operation. The encoder relies on a traditional convolutional neural network architecture that is passed the spectrogram representations of the pre- and post-gap context segments and encodes these into a single vector. The resulting encoding is then reconstructed by the decoder which uses deconvolutional layers to produce a single TF representation of the gap. The spectrogram of the gap is inserted between those of the pre- and post-gap context segments, creating one full spectrogram of the entire signal. For this full representation, the Griffin-Lim \cite{griffin-lim} algorithm is applied, obtaining the phase information necessary to return the signal back to a time-domain representation via the inverse STFT. 


\cite{audioinpainting} takes a similar approach to the one explained above, except that the authors introduce two separate models to tackle the inpainting problem. The first model deals with the problem in the audio domain, taking the raw audio with the missing sections zeroed out and a mask indicating which segments are missing. For this audio-based model, the authors make use of gated convolution layers as introduced in \cite{Gatedconv}. The second model, which operates on the spectrogram of the signal, also takes a mask indicating which portion of the signal has been lost. The same architecture is used, save for the fact that regular convolutions are used, rather than dilated ones. Similarly to the model introduced in \cite{marafioti2019context, marafioti2019audio}, the output of this spectrogram model is transformed back to the time-domain via the Griffin-Lim algorithm \cite{griffin-lim}.

\begin{table}[!b]
    \centering
    \vspace{-0.3cm}
    \begin{tabular}{c|c}
        Architecture & References \\
        \hline
         Feed-forward  & \cite{PLC2016}\\
        LSTM RNN & \cite{Lotfidereshgi2018, mywork} \\
        Autoencoder & \cite{marafioti2019context, marafioti2019audio, audioinpainting, kegler2019deep}  \\
        GAN & \cite{shi2019speech, acousticinpainting, visionaudioinpainting} \\
    \end{tabular}
    \vspace{0.1cm}
    \caption{The deep PLC models and their utilised architectures.}
    \vspace{-0.6cm}
    \label{tab:approaches}
\end{table}

Similar to the approaches introduced above, the authors of \cite{kegler2019deep} tackle PLC by way of spectrograms. They make use of a U-Net \cite{unet} architecture as their Autoencoder to produce the recovered spectrograms rather than calculating a per-pixel loss between the resulting spectrograms and the originals. The authors train the U-Net using deep feature losses by employing a VGG feature extractor network \cite{unet}, wherein the network is trained to minimise the difference between the features extracted from the reconstructed spectrogram and those of the original spectrogram. The authors introduce two variations of their architecture for different tasks: \textit{informed} inpainting, where a mask is given as input specifying which segments have been lost (similar to both approaches listed above) and \textit{blind} inpainting, where such a mask is not available. The authors did not concern themselves to transform the spectrograms into time-domain.

\subsection{GAN-based deep PLC approaches}

\begin{figure}
    \centering
    \includegraphics[width=0.5\textwidth]{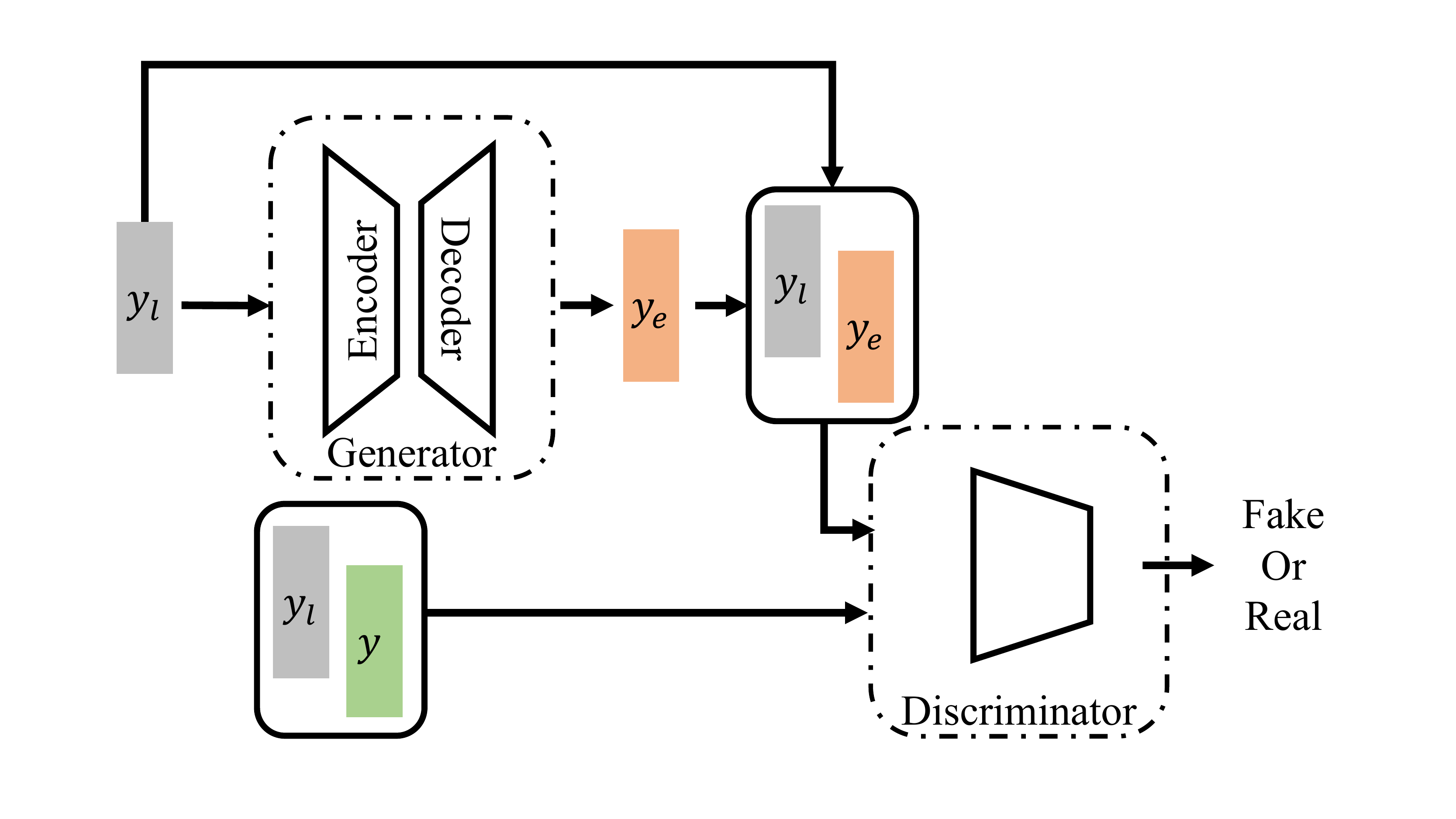}
    \caption{Framework of the SEGAN-based model \cite{shi2019speech}. $y_l$ is a lossy input audio (in grey), which is enhanced to $y_e$ (in orange) using the generator. $y$ is the ground truth signal (in green) from the training data. The discriminator tries to classify if a given enhanced signal is real (in case of an actual signal $y$) or fake (in case of the enhanced signal $y_e$). The lossy signal is concatenated to assist the discriminator. 
    }
    \vspace{-0.4cm}
    \label{fig:PLCGAN}
\end{figure}

\label{subsec:GAN}
The authors in \cite{shi2019speech} approach PLC using GANs \cite{GANS}, a class of neural networks that works by pitting two different neural networks, a generator $G$ and a discriminator $D$ against each other in a min-max adversarial fashion such that $G$ has to generate data that seemingly belongs to the distribution of the data at hand, while $D$ has to judge whether its inputs are real (actually belonging to the distribution) or fake (generated by $G$). 

Specifically, \cite{shi2019speech} is based on a GAN variant called Speech Enhancement GAN (SEGAN) that operates in the time domain by producing raw audio signals directly \cite{segan}. The SEGAN generator is constructed as an Autoencoder (as introduced in Subection \ref{subsec:autoencoders}), where the audio is encoded by using successive convolutional layers into a vector. This is concatenated with a vector of random noise and together, they are passed to the decoder which has a mirrored structure to that of the encoder. The decoder learns to recreate an enhanced version of the audio input to the encoder. In order to not lose low-level details of the input audio, the authors use skip connections between the corresponding layers of the encoder and the decoder to allow information such as phase or alignment to pass. On the other hand, given a pair of an impaired speech and its enhanced version, the discriminator $D$ is trained to classify if the enhanced speech is real (actually from the dataset) or fake (bad imitation of the dataset). The underlying principle of \cite{shi2019speech} is shown in Figure \ref{fig:PLCGAN}.

We mention two notable works due to their use of GANs as well, despite their focus not lining up totally with the focus of this paper, which is PLC for speech. The first of these works is \cite{acousticinpainting}, which focuses on audio inpainting in general using GANs, having validated their results on three different datasets of different musical instruments. They base their architecture on the Wasserstein GAN \cite{arjovsky2017wasserstein}.
The second of these works is \cite{visionaudioinpainting}, which makes use of multi-modal audio/video content. The authors make use of two independent GANs (that share a single skip connection between one of the layers of their decoder sub-networks). The first just takes the audio as input in its Mel-Spectrogram form and learns to reconstruct the spectrogram without the missing segment. The second takes as input the corrupted Mel-spectrogram, the corresponding video, motion flows extracted from the video, and the clean audio spectrogram. The components are passed to encoders whose outputs are passed to a single decoder which learns to reconstruct the clean spectrogram. Together, the outputs of both decoders are passed to a WaveNet decoder which transforms the spectrograms to time-domain audio signals.

\section{Discussion and Future Avenues}
\label{sec:challenges}

Finding a relation between Packet Loss Concealment (PLC) and other problems like Speech Enhancement (SE) and image inpainting would give a new perspective to PLC. Utilising techniques from those two related problems could help in future PLC approaches. Furthermore, a future challenge for PLC is to adapt it simultaneously with SE. 
 
\subsection{Speech Enhancement and PLC}

SE forms a similar class of problems, for example, noise is a commonly investigated aspect of them. For that, modern generative approaches like SEGAN \cite{segan} and iSEGAN \cite{iSEGAN} and attention-based approaches like \cite{attention} are used. Most of these models, however, could be also used to address packet-loss instead. Even though \cite{shi2019speech} employed this idea, still more SE ideas could be used since PLC is not explored as commonly.

More importantly, packet-loss does not always occur separately without any other speech deformations, but a variety of other aspects could occur simultaneously with packet-loss \cite{qos}. Admittedly, 
though 
isolating the packet-loss problem might lead to better solutions for it, it still limits the scope of solving other more general SE problems. Future PLC methods could attempt to overcome this by being compatible with other SE methods, or by developing more universal approaches that attempt to address issues simultaneously -- an attempt was done in \cite{mack2019deep}.

\subsection{Image inpainting}

One of the formulations of the image inpainting is solving the problem where there is a lost part of a given image that needs to be recovered. This problem is relevant to PLC in two ways. First, visual representations of the audio combined with image inpainting models can be used for packet-loss or other SE problems, as it was similarly modelled in  \cite{mack2019deep, acousticinpainting, audioinpainting, kegler2019deep, marafioti2019context}.
The second relation is that, the models used for image inpainting are mostly convolution-based models, which opens the possibility for analogous architectures that deal with audio as 1-dimensional images, by using 1D convolution/pooling/deconvolution layers instead of their 2D equivalents. \cite{wavegan} shows an example of this analogy.
These two relations open new possibilities for exploring more image inpainting solutions to solve packet-loss.

Latent representation models adapting an encoder-decoder framework have shown promising results in dealing with image inpainting. In \cite{latentManifold}, there is a sophisticated encoder mechanism that constructs high dimensional latent representations which improves the performance of image reconstruction for various problems. \cite{highresinpaint} uses a method that adapts convolution-based Autoencoders for high resolution inpainting. 
Moreover, a fast light-weight method for deep image inpainting is discussed in \cite{pepsi}. All of these are very recent improvement results for image inpainting which could be adapted for PLC.




\section{Summary}
\label{sec:conclusion}

In this mini-survey, we reviewed - to the best of our knowledge - all the deep learning approaches for Packet Loss Concealment (PLC) up to now, as listed in Table \ref{tab:approaches}. They were mainly divided into approaches that could be applied step-wise in real-time as post-processing for small packets or approaches that process larger segments offline. The real-time approaches are using a history buffer of preceding packets and use them for concealing lost packets using recurrent or feed-forward neural networks. The other approaches are using more sophisticated models like Generative Adversarial Networks (GANs) or Autoencoders, given a big segment including some lost packets, and using the whole context of non-lost packets it conceals the lost ones.

To give some background to PLC, we very briefly exposed some of the classical techniques for PLC like Linear Predictive Coding (LPC) and others. We also reviewed how packet-loss is modelled in realistic settings using Markov models. Moreover, evaluation mechanisms for PLC techniques and Speech Enhancement techniques were reviewed. 

Finally, we addressed future challenges for the packet-loss problem in the Speech Enhancement (SE) context, where future techniques should try to use already existing SE models for PLC or combine them to solve several issues simultaneously. In addition to this, we shed light on relevant modellings of the packet-loss problem in relation to image inpainting, and how recent and effective image inpainting techniques could be employed for PLC in two possible ways. Future works can address such transfer of methods to the audio domain.


\bibliographystyle{IEEEtrans}
\bibliography{ms}




\clearpage

\end{document}